\documentclass{article}

\usepackage{arxiv}

\usepackage[utf8]{inputenc} % allow utf-8 input
\usepackage[T1]{fontenc}    % use 8-bit T1 fonts
\usepackage{hyperref}       % hyperlinks
\usepackage{url}            % simple URL typesetting
\usepackage{booktabs}       % professional-quality tables
\usepackage{amsfonts}       % blackboard math symbols
\usepackage{nicefrac}       % compact symbols for 1/2, etc.
\usepackage{microtype}      % microtypography
\usepackage{amsmath}
\usepackage{graphicx}
\usepackage{multirow, bigdelim}
\usepackage{color}
\newcommand{\rmse}{$\overline{\textnormal{RMSE}}$}\;
\usepackage[numbers, sort, comma, square]{natbib}

\title{Reverse engineering neural networks from many partial recordings}

\author{
  Elahe Arani\\
  Biophysics Department\\
  Donders Institute for Brain, Cognition, and Behaviour\\
  Radboud University\\
  6525AJ Nijmegen, The Netherlands\\
  \texttt{elahe.arani@science.ru.nl}\\
  \And
  Sofia Triantafillou\\
  Department of Bioengineering\\
  University of Pennsylvania\\
  Philadelphia, PA, USA\\
  \texttt{sofi.triantafyllou@northwestern.edu}\\
  \And
  Konrad P. Kording\\
  Department of Bioengineering, and\\
  Department of Neuroscience\\
  University of Pennsylvania\\
  Philadelphia, PA, USA\\
  \texttt{kording@upenn.edu}\\
}

\begin{document}

\maketitle

\begin{abstract}
Much of neuroscience aims at reverse engineering the brain, but we only record a small number of neurons at a time. We do not currently know if reverse engineering the brain requires us to simultaneously record most neurons or if multiple recordings from smaller subsets suffice. This is made even more important by the development of novel techniques that allow recording from selected subsets of neurons, e.g. using optical techniques. To get at this question, we analyze a neural network, trained on the MNIST dataset, using only partial recordings and characterize the dependency of the quality of our reverse engineering on the number of simultaneously recorded "neurons". We find that reverse engineering of the nonlinear neural network is meaningfully possible if a sufficiently large number of neurons is simultaneously recorded but that this number can be considerably smaller than the number of neurons. Moreover, recording many times from small random subsets of neurons yields surprisingly good performance. Application in neuroscience suggests to approximate the I/O function of an actual neural system, we need to record from a much larger number of neurons. The kind of scaling analysis we perform here can, and arguably should be used to calibrate approaches that can dramatically scale up the size of recorded data sets in neuroscience.
\end{abstract}

\section{Introduction}
Reverse engineering (RE) the brain plays an important role in neuroscience. To do so, neuroscientists record the activity of neurons and try to reverse engineer their relationship using machine learning (ML) and statistical methods (e.g. see \citet{pillow2008spatio,o2015computational}).
%For example, \citet{pillow2008spatio} describe the connectivity of neurons in sensory area, \citet{o2015computational} discuss how to build models from large neural activity datasets. 
However, current technology allows us to record only a small subset of the neurons that participate in the solution of even simple tasks. Neuroscience, at least in the near future, needs to rely on RE the brain using partial recordings.

Even though it is not possible to simultaneously record all neurons involved in a task, it is possible to make multiple partial recordings potentially with different observed neurons in each recording. For example, multi-electrode measurements \cite{ballini20131024}, 2-photon calcium imaging \cite{kerr2008imaging,grienberger2012imaging} and optogenetics \cite{deisseroth2011optogenetics}, allow us to simultaneously observe a subsets of neurons. The optical methods even allow us to select which exact set of neurons we want to record. Therefore, it is natural to question if and how much integrating multiple partial recordings can compensate for the partial observation.

Based on these technological advances, a few studies have recently focused on dealing with sub-sampled observations of neural activity. 
\citet{pillow2007neural} extend the linear-nonlinear Poisson (LNP) framework to include the activity of unmeasured (hidden) neurons to estimate connectivity patterns among observed and unobserved neurons. However, the method is limited to small numbers of unobserved neurons.
\citet{wohrer2010linear} propose recovering the full noise correlation matrix from partial electrophysiology recordings, based on fully observed signal correlations. 
\citet{turaga2013inferring} use a latent dynamical system model to combine two non-simultaneously recorded but strongly interacting populations of neurons into one model, and show that combining partial observations improves the prediction of neural activity.
These studies show that there is a growing interest in combining partial observations to improve our understanding of the brain.
 
Quantifying the added benefit of combining partial information to elucidate neural function is hard, particularly in the absence of known ground truth. Recently, \citet{jonas2015automatic,jonas2017could} advocate using man-made systems, where the ground truth is fully accessible, to evaluate methods used for studying biological systems and identify their strengths and caveats. Based on this idea, in this paper, we explore how well we can reverse engineer the neural activation in an artificial neural network (ANN) based on partial observations of its hidden neurons. We see this as a step forward in answering the following questions: Is it meaningful to try and reconstruct a complex neural circuit when we only partially observe it? What is the best way of dealing with the missing information in partial recordings? Which of the ML methods is more successful for the task? And finally, what is the best strategy for combining partial observations, given that the sampling capacity is technically limited? 

In addition, in light of the recent successes of ANNs in a wide range of ML problems, reverse-engineering an ANN has become an interesting problem in its own right. In particular, reconstructing part of a networks' input-output functions based on partial observations is related to the redundancy of the specific network. Redundancy in ANNs has already been discussed from different angles: for example, inspired by biological networks, \citet{Izui1990} show that redundant neural networks are more accurate, faster and more stable. \citet{denil2013predicting} show that there is significant redundancy in the parameterization of several deep learning models and that based on a few parameters of the model it is possible to accurately predict the remaining values. From an ML perspective, this redundancy is interesting because it allows the compression of the network. \citet{Cheng2015ICCV} exploit the redundancy in ANNs to reduce memory footprint. Techniques based on knowledge distillation \cite{hinton2014} compress the knowledge of a network into a more compact model, which is trained to predict the soft outputs of the larger model. Being able to understand neural networks based on "recordings" is important both in biology and in machine learning.

There are many definitions of "understanding" but one near universally accepted aspect of understanding is the ability to make predictions. One simple way of quantifying this is the root mean squared error (RMSE) of the input/output (I/O) prediction. For the approximation of the I/O function of a system, ML provides a good number of standard approaches to supervised learning. Popular techniques include XGBoost and ANNs. In this view, being able to fit an ML algorithm to a system's I/O function that allows low RMSE is the definition of understanding the system.

In this work, we train ANNs on the MNIST dataset. We use these ANNs as the ground-truth networks, and try to estimate their intermediate I/O functions using partial observations of the hidden layers as input. This resembles reverse engineering in neuroscience, where subsets of neurons in different parts of the brain are measured, and ML methods are used to estimate the I/O relationship between them.

\section{Problem Formulation}
Reverse engineering in neuroscience often aims to understand the functionality or I/O function of areas in the brain. However, brain measurements are noisy and limited by the recording technologies, and the ground truth is not known. Here, we use standard ANNs trained on a classical hand-written digit dataset (MNIST) as the ground truth systems. We use 3000 samples from the MNIST training data to train our ANNs, while the rest is used to produce partial recordings data, described below. Assuming a network of $L+1$ layers (with $l=0$ being the input layer and $l=L$ the output layer), with $N_l$ neurons on each layer, the activity of the neurons in the output layer $L$ is a function of the activity of any previous layer $l$:
\begin{equation}\{O_i^L\}_{i=1}^{N_L} = f(\{O_j^l\}_{j=1}^{N_l}),\end{equation}
where $O_n^l$ is used to denote the output of neuron $n$ in layer $l$. In this work, we train two different ANNs on the MNIST data. Details on network architectures and their performance are shown in Table \ref{tab:nns}. As these ANNs are just our ground truth system, the exact details should not matter much. We now have the ground truth system which we can try to reverse engineer to check how well various methods work based on multiple partial recordings.

\begin{table}
\centering
\caption{\label{tab:nns}The ground-truth neural networks.}
\resizebox{\textwidth}{!}
{\begin{tabular}{|cccccc|} \hline
&\#Hidden&\# Neurons in&\# Neurons in&Accuracy on&Accuracy on\\ 
&Layers&Hidden Layers& Output Layer& Training Data& Test Data\\\hline  
NN&4&128, 128, 128, 128&10&99.62&97.36\\ 
DNN&7&512, 256, 256, 128, 128, 128, 128&10&99.55&97.78\\ \hline 
\end{tabular}}
\end{table}

We want to reverse engineer the activity of the output neurons based on observations of subsets of hidden units, which are a stand-in of potentially recorded neurons in the brain. We will assume that at any given point of time, we can only record a subset of neurons, and will thus have considerable missing information. We want to quantify how well RE is possible based on an ML technique, the approach to deal with the missing data (unrecorded neurons), the number of recorded neurons, and the noise level. 

\begin{figure}
\begin{tabular}{cccc}
\includegraphics[width=0.22\columnwidth]{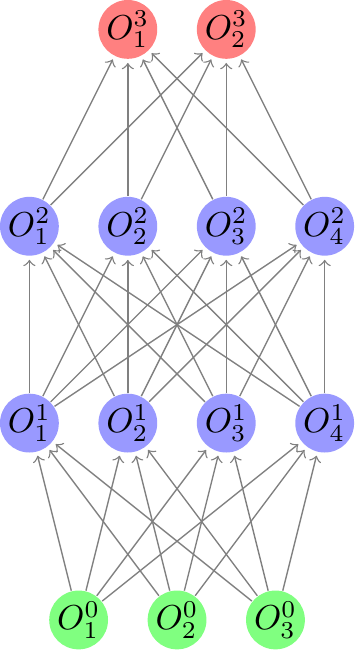}&
\includegraphics[width=0.22\columnwidth]{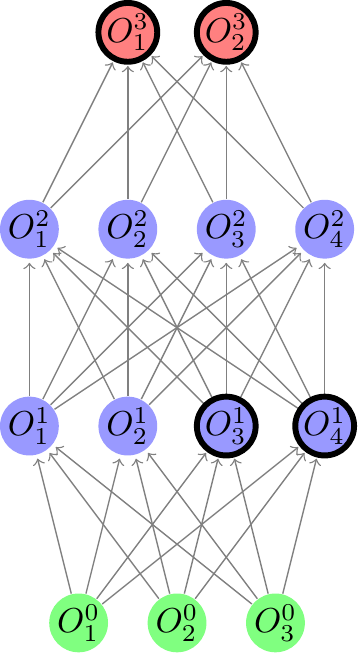}&
\includegraphics[width=0.22\columnwidth]{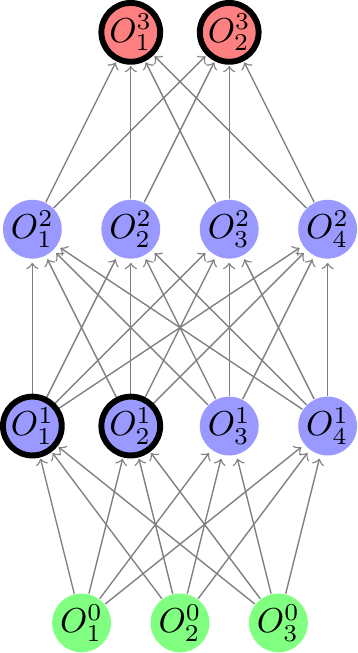}&
\includegraphics[width=0.22\columnwidth]{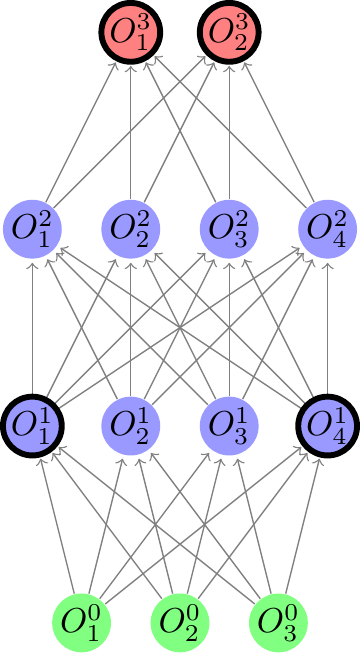}\\
(a) Trained NN&(b) Subset 1&(c) Subset 2&(d) Subset 3\\
\end{tabular}
\caption{\label{fig:nn}An example of three recording subsets. (a) The ground-truth neural network. (b,c,d): An example of three ($K=3$) settings. In each setting, the output neurons ($O_1^3,O_2^3$) and two of the neurons of layer $1$ (a:$O_3^1,O_4^1$, b:$O_1^1,O_2^1$, c:$O_1^1,O_4^1$) are observed. Observed neurons are outlined in black.} \end{figure}

To address these issues, we will simulate partial recordings from the trained ANNs and use ML to reverse engineer the hidden-unit to output-unit mapping. We assume that there is a sequence of settings $k$. For each setting, only a random subset of neurons is observable. We can think of each k, as one setting of the electrical or optical recording apparatus. For each setting, we thus observe the activity of a different subset of the neurons $\{O_j^l\}_{j\in Obs_k}$, while the rest are unobserved. An example of partial data is shown in Fig. \ref{fig:nn}. This way, we produce simulated training sets that approximate what we may record partially from a real brain.

Based on the data collected from the (partial) recordings, we want to estimate the activity of the output neurons as a function of all the observed neurons in layer $l$. More specifically, we want to estimate:
\begin{equation}\label{eq:pr_model}\{\hat O_i^L\}_{i=1}^{N_L} = \hat f(\{O_j^l\}_{j\in\cup_k{Obs_k}}).\end{equation}
This problem is not trivial because the partial recordings produce many missing values corresponding to unobserved neurons in each recording. The ML methods thus need to gracefully deal with missing data. This allows us to quantify how well we can reconstruct the I/O function of the ANNs based on different numbers and nature of partial recordings.

\section{The Effect of Partial Recordings on Reconstruction Quality}

To reverse engineer a system from partial data, we can use an ML method for regression, but this method needs to deal with massive amounts of missing data. The missing data have a specific structure: in each recording, only a subset of variables (neurons) are observed. The final model in Eq.\ref{eq:pr_model} is defined over the union of all variables that have been observed in at least one of the recordings. The ratio of missing over observed data is higher for recordings with fewer observed neurons per recording. 

Missing data is a common problem in statistics, and the way it is dealt with can affect the results, depending on the missingness mechanism. In the simplest case, where data are missing independently of observed and unobserved data, the data are said to be Missing Completely At Random (MCAR) \cite{Rubin1976}. This is the case for the partial data, since the inclusion of a neuron in a recording is based on the sampling design, and does not depend on the activity of the neurons, observed or unobserved. When data are MCAR, excluding the samples with missing values do not bias the results. Arguably, neuroscience is not MCAR. For example, as small neurons are more likely to be unobserved than big neurons because both their optical and electrical signatures are harder to pick up. However, discarding every sample that includes a missing value is not possible here, since every recording has some missing values. Thus, we must select a strategy for dealing with the missing data.

When discarding missing data is not an option, the most common remedy is to impute the missing values in a preprocessing step. Mean value imputation (MVI) is common, where every missing value for a variable is replaced with its sample mean. More complex approaches in the area of multiple imputation \cite{rubin1996multiple} fill in the missing values by assuming a model for the missing values that can be estimated from observed values. The estimated model is then used to impute the missing values. For example, Soft-Impute (SI) iteratively uses a soft-threshold singular value decomposition (SVD) to replace the missing values \cite{mazumder2010spectral}. 
A few approaches in statistics try to deal with combining data with overlapping observed variables (similar to partial recordings data). In the field of causal discovery, several methods can learn causal structure from multiple partial observation, based on the assumption of a single underlying causal process \cite{triantafillou2015constraint}. However, these methods are used for learning causal networks and are not applicable in this setting. The field of statistical matching \cite{dorazio2006} proposes some methods for missing data imputation or for inferring the parameters of the joint probability distribution of the union of observed variables without explicitly imputing the missing values. However, these methods require at least some variables measured in common and are therefore not directly applicable in the general cases. Thus, here we use MVI and SI to complete the missing values in our partial recordings data.

After the missing value imputation, the complete data set can be used to train an ML method to estimate Eq.\ref{tab:nns}. We use ANNs and XGBoost in this work. XGBoost \cite{chen2016xgboost} also has a built-in approach for handling missing data without explicitly imputing the missing values, by adding a default direction to each tree node. When a value is missing, the instance is classified in this default direction, which is estimated from the data. Thus, ANNs have to be coupled with an imputation strategy, while XGBoost can also be used directly on the partial data. 

An alternative approach to data imputation is training a separate estimator for each recorded subset and combining their predictions. In this case, the estimator in Eq.\ref{eq:pr_model} can be computed as the average of $K$ marginal estimators: 
\begin{equation}\label{eq:pr_models_mp} 
\hat f(\{O_j^l\}_{j\in\cup_k{Obs_k}}) = \frac{1}{K}\sum_k \hat f_k(\{O_j^l\}_{j\in Obs_k})
\end{equation}

The advantage of marginal estimators is that each $\hat f_k$ can be identified with any appropriate ML method without preprocessing, in parallel. However, each marginal estimator is then estimated based on fewer samples. 

In the next section, we study the quality of reverse engineering based on multiple partial recordings, and how it depends on different factors, using simulated data from the ground truth networks.

\section{Experiments}
Here, we use simulated experiments to examine how the quality of reconstruction based on partial recordings depends on: (a) the data imputation method, (b) the size of the ground truth network, (c) the number of different recorded subsets and (d) the noise level. We use the networks in Table \ref{tab:nns} as our ground truth networks, and simulate observations by sampling with replacement from the second half of the MNIST training data that were not used for training the ground truth networks. We typically simulate $K=10$ partial recorded subsets. This means that we consider $K$ subsets of neurons to record from, each chosen randomly (and thus often overlapping), which corresponds to switching the subset of neurons recorded $K-1$ times. In each subset, all output neurons and $N$ neurons from a given hidden layer are observed. We then train an ML method to predict the values of the output neurons based on the partial data of the given hidden layer. We calculate the average (over neurons) of the Root Mean Squared Error (denoted by \rmse) over all output neurons, then we report the mean and the standard deviation of \rmse\ over $5$ iterations as a measure of the quality of our RE. We can thus ask how all the mentioned factors relate to the quality of reverse engineering.

The two popular methods we will look at are XGBoost and ANNs as they represent probably the two most popular approaches in supervised regression. More specifically, We use them coupled with different imputation techniques (MVI and SI), as well as with combining separate marginal models (denoted by MP for mean prediction). In addition, we used XGBoost with the built-in approach for handling missing data (denoted by XBG-G). For each partial data set, we trained an independent model with XGBoost for every output neuron, and a single ANN model with 10 output neurons. XGBoost was used with the following parameters: $\eta=0.5$ and $max\_depth=6$. We used an ANN with three hidden layers ($16$ neurons in each, ReLU activation function) and applied early-stopping based on the loss value. Having implementations of all methods allows us to ask how they perform at reverse engineering the ground truth ANNs.

\begin{figure} \centering
\includegraphics[trim=.5cm 0cm 1cm 1cm, clip, width=.55\columnwidth]{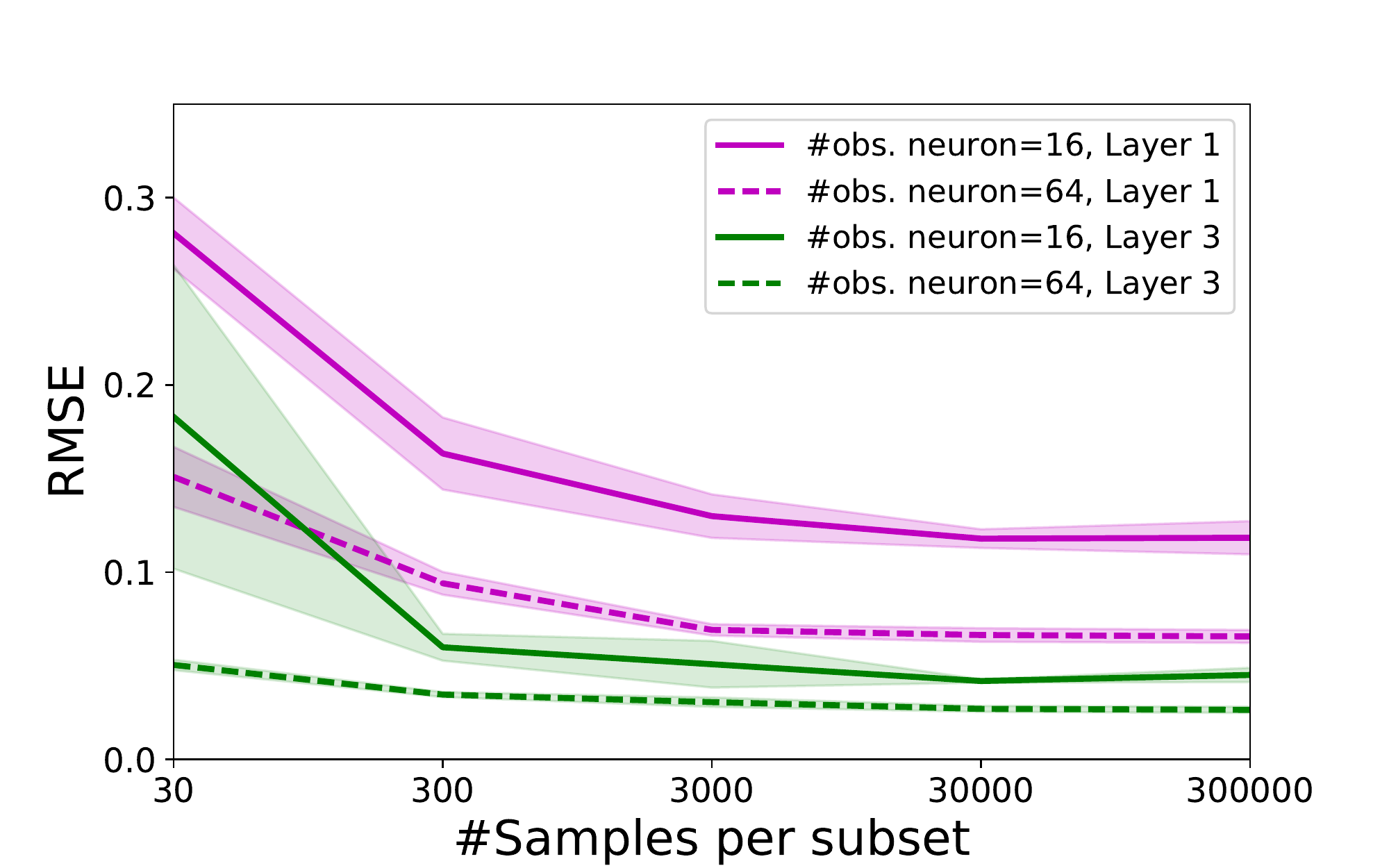}\\
\caption{\label{fig:sample_size}\textbf{Unsurprisingly, having more data leads to better reconstructions.} 
\rmse\ over $5$ iterations as a function of sample size per setting from the first and the third hidden layer of the NN, using ANN-MVI.} \end{figure}

We should expect that having more data helps us better predict output responses. Indeed, prediction quality improves but saturates at a level that depends on the number of simultaneously recorded neurons (Fig. \ref{fig:sample_size}). The same qualitative behavior was seen for the methods we do not show in the figure which we suppressed for clarity. We may expect and observe that the saturation level that can be reached may depend strongly on the number of neurons in the system, the manifold of stimuli used, and the nonlinearity of the system.

To make comparisons meaningful, we will for the rest of the paper, only compare situations where the same total amount of information is acquired, i.e. each training set will have the same product of the number of observed neurons and the number of observations. This is meant to mimic the constraints of real neuroscientific experiments where fixed information throughput limits the overall recorded data set size. You can either opt to record more neurons (larger subset sizes) for fewer samples, or fewer neurons for more samples. 

\begin{figure} \centering
\includegraphics[trim=3.55cm 0cm 3.8cm 0cm, clip, width=\columnwidth]{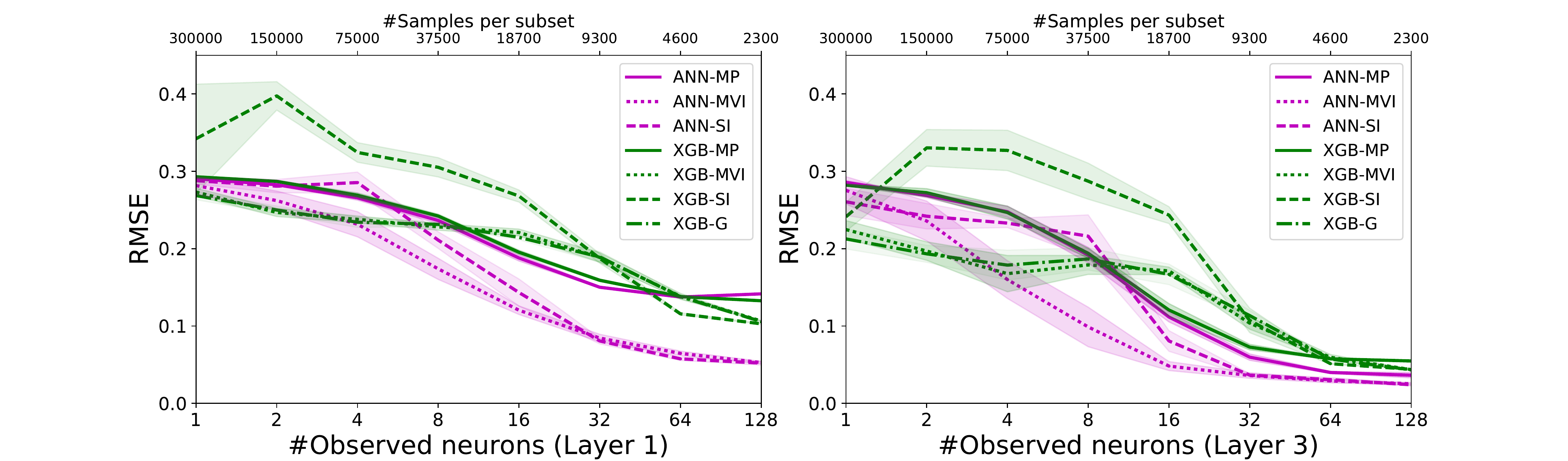}\\
\caption{\label{fig:RMSEs}\textbf{Neural networks with mean value imputation work best at reverse engineering an 
ANN.} 
\rmse\ as a function of the subset size from the first (left) and third (right) layer of the ground truth NN.} \end{figure}

We want to know which method is better at dealing with many partial recordings. We thus measure the quality of RE for different methods and a different number of observed neurons in each recording (Fig. \ref{fig:RMSEs}). Both layers have similar performance. The error decreases as the number of observed neurons increases, although the total number of non-missing data points remains nearly the same. For both layers, ANNs with MVI perform better for almost all settings.

\begin{figure} \centering
\includegraphics[trim=3.55cm .2cm 3.8cm .2cm, clip, width=1\columnwidth]{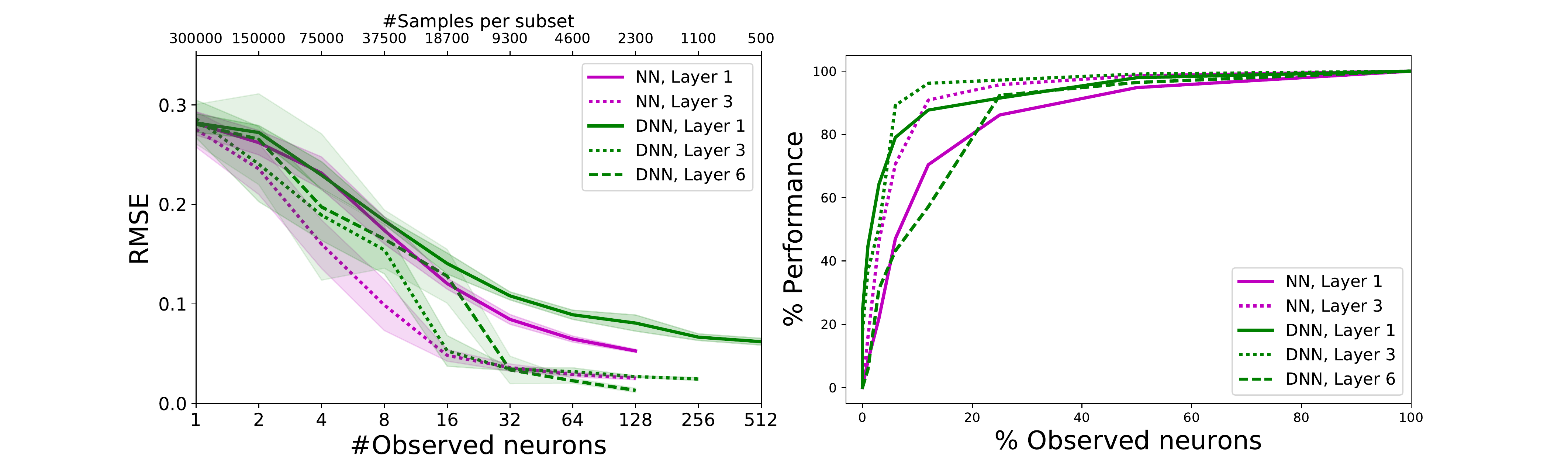}
\caption{\label{fig:NN_DNN} \textbf{Simultaneously recording more neurons helps reverse engineering for a fixed number of samples, just as well for shallow and deep networks.}
Left: \rmse\ for RE as a function of observed subset size using ANN-MVI in layers $1, 3$ of the NN and in layers $1,3,6$ of the DNN. Right: Performance (as a percentage of maximum decrease in \rmse) vs percentage of observed neurons for each layer.} \end{figure}

How successfully can we reverse engineer a deeper and wider neural network? And, how many neurons must we observe to have reasonably good error rate? To address these questions, we perform the experiments using ANN-MVI with data simulated from a deeper neural network (DNN) as the ground truth network; to compare NN and DNN, see Table \ref{tab:nns}. Results in Fig. \ref{fig:NN_DNN} show similar trends in both NN and DNN data: increasing the number of observed neurons improves performance. Moreover, for wider layers, a smaller ratio of neurons is required for adequate performance. In all layers, by observing roughly twenty percent of the neurons, RE achieves nearly the same error rate as with a fully observed layer. Therefore, ANN with MVI may be a reasonable choice for deep neural networks.

\begin{figure} \centering
\hspace*{-.55cm}
\includegraphics[trim=4.6cm 1.1cm 0cm 1.7cm, clip, width=1.19\columnwidth]{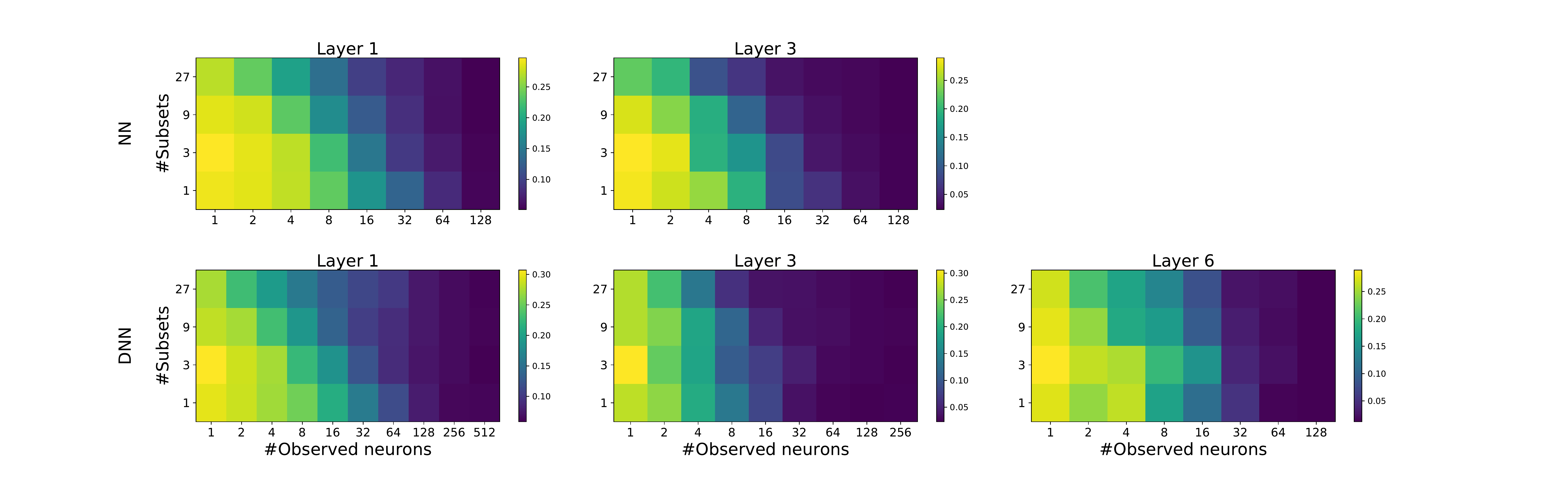}
\caption{\label{fig:par_rec} \textbf{Sequentially recording from many subsets of neurons can compensate for recording only small subsets at a time.}
\rmse\ for RE using ANN-MVI for a different number of sequentially recorded subsets $K$, keeping the total number observed values approximately constant. Results are shown for layers $1,3$ of the NN and $1,3,6$ of the DNN.} \end{figure}

Results so far show that increasing the number of the observed neurons at the expense of presenting fewer stimuli improves RE for a fixed number of selections of subsets of neurons. However, it is possible that increasing the number of selected subsets can compensate for few simultaneously recorded neurons. Choosing many different subsets is often experimentally feasible, e.g. by focusing a laser on different subsets of neurons on a different plane. To test the trade-off between the number of subsets and number of recorded neurons in each, we simulated partial data for varying $K$, and a different number of observed neurons per subset. RE with a single selection of neurons is not meaningfully possible for small numbers of recorded neurons (Fig. \ref{fig:par_rec}). However, increasing the number of selected subsets improves the quality of the reverse engineering; this improvement is more noticeable when fewer neurons are simultaneously recorded. While the changes in performance are not always monotonic, in general, combining more subsets can compensate for the lack of full observations. Thus, combining more recordings of small subsets, we can achieve similar performance to combining fewer recordings of larger subsets.

\begin{figure} \centering
\includegraphics[trim=3.55cm .2cm 3.8cm 0cm, clip, width=1\columnwidth]{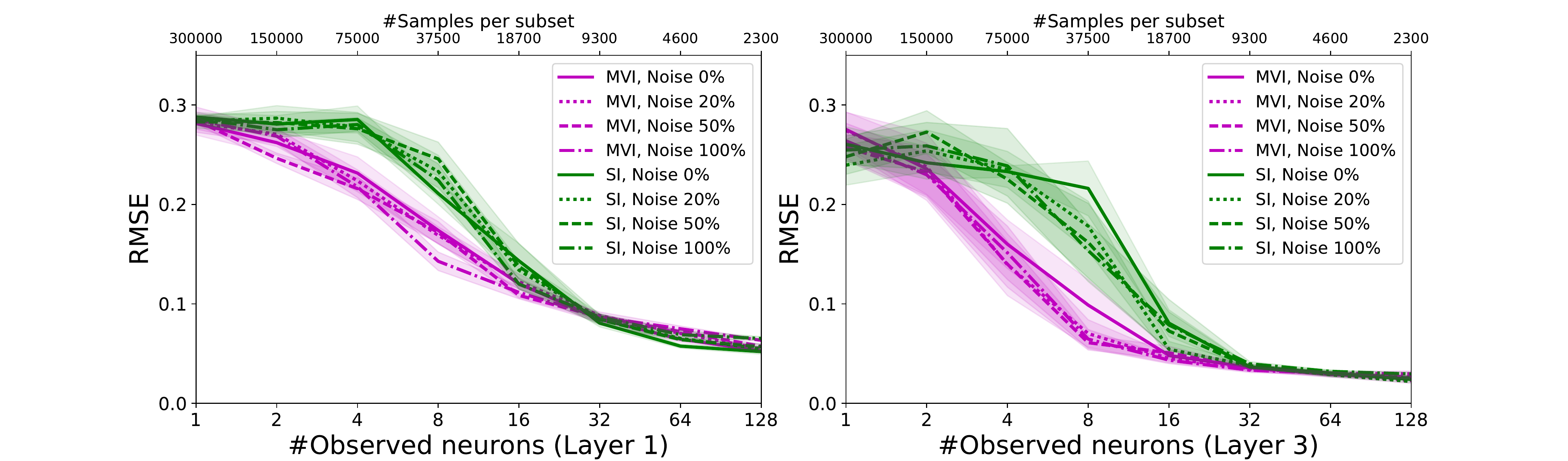}
\caption{\label{fig:noise}\textbf{Noise has little effect but mean value imputation beats soft imputation.}
\rmse\ for RE as a function of the observed subset size using ANN-MVI and ANN-SI with various additive noise levels.} \end{figure}

The recordings of neural networks include several sources of noise. The effective noise in the recordings has a component that comes from measurement, e.g. Johnson noise in the electrode, and a component that comes from the brain itself (e.g. Poisson noise in spiking). To study the effect of noise on RE, we add signal-dependent noise to the partial recordings. For each sample, we add zero-mean Gaussian noise with variance that is proportional to the activity of the neuron in the given sample. We set the noise variance to $20, 50$ or $100$ percentage of the neurons' activity value. Here, $100\%$ corresponds to the level expected for a Poisson process. We add the noise after simulating the partial recordings and use ANNs to estimate the I/O function. To check which of the imputation techniques can handle the noise better, we consider both MVI and SI imputation techniques. MVI still outperforms SI for partial data with additive Gaussian noise, but both methods are robust to all levels of additive signal-dependent noise (Fig. \ref{fig:noise}).

\begin{figure} \centering
\includegraphics[trim=0cm 0cm 1cm 1cm, clip, width=.55\columnwidth]{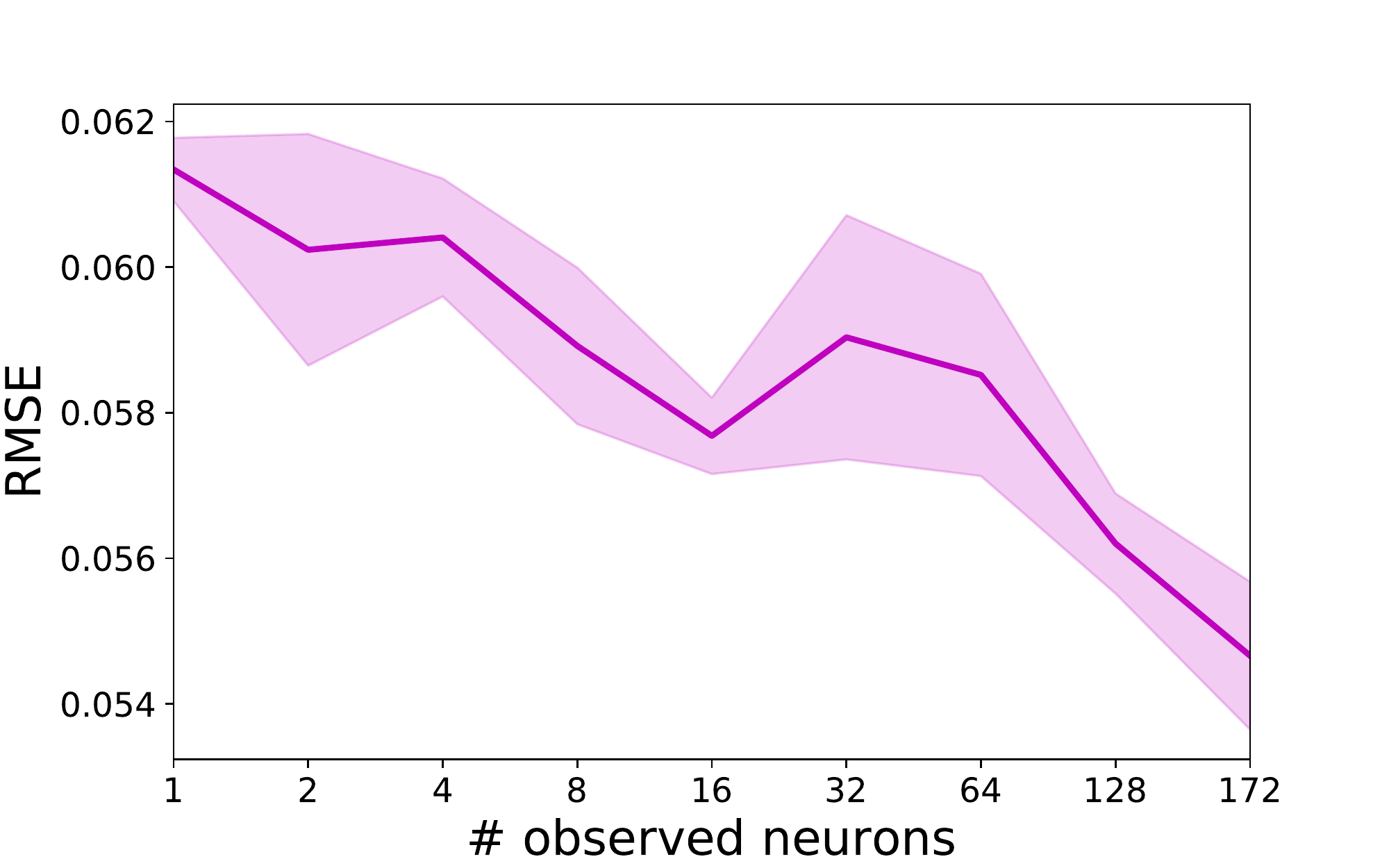}
\caption{\label{fig:spiking}\textbf{Recording from larger subsets of neurons in PMd helps for predicting hand kinematics.}
\rmse\ for RE as a function of the observed subset size using XGBoost-MVI on spiking data.} \end{figure}

To check if our scaling analysis is applicable to current neuroscience datasets, we used spike count data (50ms bins for 172 neurons) from cortical area PMd \citet{stevenson2011statistical} to predict hand velocity of a monkey, with the same setting as simulations. We then reverse engineer the I/O function using XGBoost with MVI. Similar to the simulations, performance improves with larger subset sizes (Fig. \ref{fig:spiking}). However, this behavior does not even get close to reaching saturation. This is expected because in this dataset we only record from order hundreds of millions of neurons involved in a hard task (life). Observing more neurons simultaneously helps, but not all that much. However, we do believe that recordings from considerably more neurons are necessary to make this analysis meaningful.

\section{Discussion}
Modern techniques in neuroscience allow us to simultaneously record from many neurons. Motivated by this, we investigated how well we can reverse engineer neural activity in ANNs from multiple partial observations. Surprisingly, MVI - the simplest and easiest imputation technique - achieves a lower error rate for RE a system with multiple partial recordings even compared to a more complex and computationally expensive method, like Soft Impute. Also, we found that increasing the number of simultaneously observed neurons increases the quality of our RE. However, many recordings from small subsets of neurons can well approximate what we get if we simultaneously record from all neurons. In real neural data, observing more neurons simultaneously also improves the quality of RE, however, since we can only observe a tiny part of the actual system, this behavior does not reach a saturated point.
Thus, combining multiple partial recordings may improve our understanding I/O functions in the brain.

Obviously, an artificial neural network is not a brain. Even though ANNs are inspired by neuroscience, brains are biophysically more complex and probabilistic. Biological neurons can perform on many different time scales and are more heterogeneous in many ways. In the brain, the communication takes place through spikes, whereas it happens via abstract rates in neural networks. Moreover, in our specific work here, we dealt with a tiny ANN. Our analysis might give categorically different answers if we could apply it to the real brain.

The architecture of our ground truth ANNs is obviously idiosyncratic: The networks we use are simple feed-forward neural networks with no lateral or recurrent connections. More complex architectures, typical in the brain, may produce different behavior. In the future, it would be interesting to extend our analysis to recurrent neural networks and to networks with lateral connections.

In all our simulations, the ground truth ANNs used were trained on the MNIST dataset, which may have an atypical structure. After all, for each of the digits, there are a small number of universally accepted generative models that are transmitted culturally at places called Kindergarten. This means that the manifold of produced digits will have a far simpler structure than typical manifolds of things in the real world. Moreover, this structure makes the MNIST problem easy from an ML perspective and may lead to low dimensional solutions that may lead our technique to produce a low estimate for the number of neurons that need to be simultaneously recorded.

This work presents a test case where artificial systems are used to evaluate and guide computational neuroscience. Using ANNs as a stand-in for an actual neural circuit, we found that reverse engineering neural function based on a small fixed subset of recorded neurons is not possible, even for a low dimensional task. Using real spiking data we found some evidence that such behavior may also be real for the brain. This may suggest that the number of neurons that need to be simultaneously recorded may be larger for more complicated tasks. We, therefore, believe that a similar scaling analysis should be standard for reverse engineering in neuroscience applications.

Modern neuroscientific techniques allow running similar analyses to the one we presented here on real brains. Using Ca2+ imaging \cite{kerr2008imaging,grienberger2012imaging} along with modern optical targeting techniques allows trading of the number of recorded neurons with the noise level. It also allows simultaneous recording from the input parts of a system and the outputs. As such, the neural network that we used here, could be readily replaced with a real sample of brain tissue. Doing scaling analyses on reverse engineering approaches is the only way of knowing how recording techniques need to be optimized.

% \section*{Acknowledgement}
% E.A. was supported by HealthPAC (FP7-PEOPLE-2013-ITN).

%\section*{References}
% \bibliographystyle{unsrt}  
\bibliographystyle{unsrtnat}
\bibliography{bibliography}
\end{document}